%
\input phyzzx.tex
\tolerance=1000
\voffset=-0.3cm
\hoffset=1.0cm
\sequentialequations
\def\rl{\rightline}

\def\t1{{\tilde 1}}

\def\NPB#1#2#3{Nucl. Phys. B {\bf#1} (19#2) #3}

\REF\DB{J. Polchinski, S. Chaudhuri and C. Johnson, hep-th/9602052 and references therein.}
\REF\SV{A. Strominger and C. Vafa, hep-th/9601029; C. Callan and J. Maldacena, hep-th/9602043.}
\REF\HMS{G. Horowitz, J. Maldacena and A. Strominger, hep-th/9603109.}
\REF\HS{G. Horowitz and A. Strominger, hep-th/9602051.}
\REF\FOUR{J. Maldacena and A. Strominger, hep-th/9603060; C. Johnson, R. Khuri and R. Myers, hep-th/9603061; G. Horowitz, D. Lowe and J. Maldacena, hep-th/9603195.}
\REF\ROT{J. Breckenridge, R. Myers, A. Peet and C. Vafa, hep-th/9602065; J. Breckenridge, D. Lowe, R. Myers, A. Peet, A. Strominger and C. Vafa, hep-th/9603078.}
\REF\DM{S. Das and S. Mathur, hep-th/9606185; hep-th/9607149.}
\REF\MS{J. Maldacena and L. Susskind, hep-th/9604042.}
\REF\SEN{A. Sen, \NPB{440}{95}{421}, hep-th/9411187; Mod. Phys. Lett. {\bf A 10} (1995) 2081, hep-th/9504147; A. Peet, hep-th/9506200.}
\REF\LEN{L. Susskind, hep-th/9309145.}
\REF\HRS{E. Halyo, A. Rajaraman and L. Susskind, hep-th/9605112.}
\REF\LW{F. Larsen and F. Wilczek, hep-th/9511064.}
\REF\LS{D. Lowe and A. Strominger, Phys. Rev. {\bf D51} (1995) 1793, hep-th/9410215.}
\REF\TSY{A. Tseytlin, Mod. Phys. Lett. {\bf A 11} (1996) 689, 9601177; hep-th/9605091; J. Russo, hep-th/9606031.}
\REF\MAL{J. Maldacena, hep-th/9605016.}
\REF\HKRS{E. Halyo, B. Kol, A. Rajaraman and L. Susskind, hep-th/9609075.}

\singlespace
\rl{SU-ITP-42}
\rl{\today}
\pagenumber=0
\normalspace
\medskip
\bigskip
\titlestyle{\bf{ Reissner--Nordstrom Black Holes and Strings with Rescaled Tension}}
\smallskip
\author{ Edi Halyo{\footnote*{e--mail address: halyo@dormouse.stanford.edu}}}
\smallskip
\centerline {Department of Physics} 
\centerline{Stanford University} 
\centerline {Stanford, CA 94305}
\smallskip
\vskip 2 cm
\titlestyle{\bf ABSTRACT}

We show that extreme and nonextreme Reissner--Nordstrom black holes in five
dimensions can be described by closed fundamental strings with two charges
in a magnetic five brane. Due to the five brane background the string oscillator
number and tension are rescaled whereas the mass remains fixed. The black hole mass 
is given by the sum of the five brane and string masses.
Its entropy however
is given by only that 
of the string with the tension rescaling taken into account. 
We also show that the emission of a 
low energy scalar from the slightly
nonextreme string matches the Hawking radiation
expected from the black hole.

\singlespace
\vskip 0.5cm
\endpage
\normalspace

\centerline{\bf 1. Introduction}

During the last year there has been remarkable progress in understanding the microscopic 
structure of Ramond--Ramond (RR) charged black holes in type II string theory by using
the D--brane technology[\DB]. These black holes are made of a collection of
different types of D--branes with the mass of the black hole[\SV,\HS]. By counting the open strings 
which end on different D--branes one can count the number of
states of the system which reproduces the black hole entropy. 
This was done for five and four dimensional extreme and nonextreme
black holes[\SV,\HMS,\FOUR] (and also for rotating ones[\ROT]). Furthermore, emission of low energy
scalars from the D--brane configuration was shown to match the Hawking radiation
expected from the corresponding black hole[\DM]. 
The picture that emerges from these works is as follows. Extreme black holes are
very well described by noninteracting BPS states of a long (D) string[\MS]. For
small deviations from extremality, the essential degrees of freedom of the black hole are
weakly interacting non BPS excitations of a long closed (D) string. 
This is also true for low energy
processes such as radiation of scalars from the black hole. Other branes present such
as five branes do not play a dynamical role but simply restrict the string oscillations to
lie inside their world--volume. Thus, the near extreme black hole seems to behave like
a weakly interacting system in one dimension which has a close resemblance to a
(D) string.

On the other hand, it was appreciated for some time that a sufficiently massive string 
state becomes a black hole since its Schwarzschild radius becomes smaller than its size
[\SEN].
Even though state counting for strings is well--known,  a simple connection between
black holes and strings has not been established. This is mainly due to the fact that naively 
the entropy of a string is proportional
to its mass whereas for the black hole $S_{BH} \sim M^2$. This apparent discrepancy
was eliminated in ref. [\LEN]. There it was shown (for Schwarzschild black holes)
that a string in its own gravitational field
(or Rindler space) has a Rindler energy which gives the correct scaling for the
black hole entropy. This 
result was later expanded for Schwarzschild black holes in all dimensions with the
correct coefficient[\HRS]. Another potential problem for the stringy description of black holes
is the fact that a string can carry at most two charges whereas a black hole can 
have more. This requires the introduction of another object into the description as
we will see below.

In this paper, we consider Neveu--Schwarz (NS) charged 
Reissner--Nordstrom (RN) black holes in five dimensions
(in type II string theory compactified on $T^5$). We show that these black holes can be
described by closed fundamental strings with two charges in a magnetic NS five brane
background. The essential degrees of freedom of the black hole are the string states. 
The extreme and nonextreme black holes are described by the BPS and non BPS states
of the string respectively. The five
brane background restricts the string oscillations to lie in its world--volume and provides the
Rindler space--time due to which the string oscillator number and 
tension are rescaled. The mass of the black hole
is given by the sum of the masses of the five brane and the string. Naively the string
does not reproduce the black hole entropy. This is not surprising since the naive string
entropy formula holds for a free string whereas we have a string in a strong gravitational
background. In order to find the entropy, we calculate the Rindler energy of the string
in the five brane background and identify $S=2\pi E_R$. 
This method
gives the correct black hole 
entropy for $c_{eff}=6$. 
We interpret this effective central charge to mean that the string oscillations are confined
into the world--volume of the five brane.
Since $E_R$ is dimensionless and has the above relation to entropy
we identify it with the (square root of the)
string oscillator number in Rindler space. Comparing this with the free
string we find that the oscillator number is rescaled due to the gravitational background.
This scaling factor is exactly what is needed to transform the naive string entropy into
that of the black hole.
On the other hand, the string mass (added to the five brane mass) 
reproduces the correct black hole mass. 
We can hold the string mass fixed under a rescaling of the oscillator number only if
we rescale $\alpha^{\prime}$ (or the string tension, $T=1/2 \pi \alpha^{\prime}$) 
by the same factor.\foot{For strings with rescaled tension in a somewhat different
context see [\LW].}
Thus, we conclude that the effect of the five brane background on the string is to rescale
its oscillator number and tension by the same factor while the string mass remains fixed.
This picture can be considered S dual to that of ref. [SV] in terms of D branes in type IIB
string theory.

Our main assumption is that all gravitational effects in the black hole are
summed up in the rescaling of the oscillator number and the string tension due to the 
five brane background. The string mass remains fixed, i.e. in the black hole 
it is equal to its value for the free string both for BPS and non BPS states. This makes 
sense for the BPS states but not for the non BPS ones which is reminiscent of
the nonextreme cases in the D--brane picture.
We have no explanation for this other than pointing out that the method works for
both extreme and nonextreme RN black holes in five dimensions. Here, 
we are also neglecting
any possible effects due to the mass of the string on itself.
This is justified for a black hole with one charge (a five brane) since the string
is a small perturbation on this background. On the other hand, for a RN black hole,
the string has two charges and weighs twice as much as the five brane. In this case, it 
is hard to understand why these effects are absent. Perhaps a  deeper understanding 
of these issues may be reached by investigating strings in Rindler space--times [\LS] and/or
by considering more fundamental descriptions of the near horizon region along the
lines of refs. [\TSY].

The paper is organized as follows. In section 2, we review the classical  RN black hole
solution in type II string theory. We define the parameters of the string picture in terms
of the parameters of the classical solution. In section 3, we first review the case of a five
dimensional black hole with one charge, i.e. a five brane. We introduce our method in 
detail for this simpler case and show that small deviations from extremality are described
by a neutral string with a rescaled tension. We then consider extreme and nonextreme
RN black holes in five dimensions. We show that by using our method their masses and
entropies are given by a string with two charges and a rescaled tension in a
five brane background. In section 4, we calculate the rate of emission of a low energy
scalar from a slightly nonextreme string and show that it matches the expected Hawking
radiation exactly. Section 5 contains a discussion of our results and our conclusions.

\bigskip
\centerline{\bf 2. The Classical Reissner--Nordstrom Black Hole in Five Dimensions}

In this section, we review the solution for the NS charged RN black hole in five
dimensions[\HMS]. The classical solution to the low energy equations of motion in type 
II string theory
compactified on $T^5$ is given by the metric $g_{\mu \nu}$, the NS antisymmetric tensor
$B_{\mu \nu}$  and the dilaton $g^2=e^{-2 \phi}$.
The RR three form, the self--dual five form and the RR scalar are set to zero. Also, the
asymptotic value of the dilaton $\phi$ is taken to be zero.
The classical five dimensional RN black hole metric is given by
$$ds^2=-f^{-2/3} \left(1-{r_0^2 \over r^2} \right)dt^2+f^{1/3} \left[\left(1-
{r_0^2 \over r^2} \right)^{-1}dr^2+r^2d\Omega_3^2 \right] \eqno(2.1)$$
where
$$f=\left(1+{{r_0^2 sinh^2 \alpha}\over r^2} \right) \left(1+{{r_0^2 sinh^2 \beta} \over r^2} \right)
\left(1+{{r_0^2 sinh^2 \gamma} \over r^2} \right) \eqno(2.2)$$
The solution is parametrized by six parameters, $\alpha, \beta,\gamma, r_0$ and
the compactified one and four volumes $2 \pi R$ and $(2 \pi)^4 V$.
The total energy of the black hole is
$$E={RV r_0^2 \over {2g^2 \alpha^{\prime 4}
}}(cosh 2\alpha+ cosh 2\beta + cosh 2 \gamma) \eqno(2.3)$$
The entropy of the black hole is found from the area of the horizon using the
Bekenstein--Hawking formula
$$S={A_H \over {4G_5}}={2 \pi RV r_0^3 \over {g^2 \alpha^{\prime 4}}} cosh \alpha~ cosh \beta~
cosh \gamma \eqno(2.4)
$$
where the ten and five dimensional Newton constants are given by 
$
G_{10}=8 \pi^6 g^2  \alpha^{\prime 4}
$ 
and 
$
G_5=G_{10} / (2\pi)^5 RV
$.
From the first law of thermodynamics and using eqs. (2.3) and (2.4) we find the
Hawking temperature
$$T_H={1 \over{2 \pi r_0 cosh \alpha~ cosh \beta~ cosh \gamma}} \eqno(2.5)$$
The RN black hole carries three NS charges
$$\eqalignno{Q_5&={r_0^2\over{2 \alpha^{\prime}}} sinh (2 \alpha) 
&(2.6a) 
 \cr
Q_1&={V r_0^2\over{2 g^2 \alpha^{\prime 3}}} sinh (2 \beta) &(2.6b)\cr               
n&={R^2V r_0^2\over{2 g^2 \alpha^{\prime 4}}} sinh (2 \gamma) &(2.6c)}$$
These are the charges of the black hole under the NS three form $H_3$, its
dual $H_7$ and Kaluza--Klein two form coming from the metric. Note that due to the
fact that we have a RN black hole the scalars in the solution, i.e. the dilaton $\phi$
and the  compactification volumes $R,V$ (which are a priori fields) are constant in space.
The RN condition is $\alpha=\beta=\gamma$
so that the three contributions to the mass of the black hole are equal.
The extreme limit is obtained by $r_0 \to 0$ and $\alpha, \beta, \gamma \to \infty$
with the charges $Q_1,Q_5,n$ fixed. The cases with less than three charges are
obtained by setting the corresponding angles to zero. 

The properties of the black hole can be written in a suggestive way if we trade
the six parameters $\alpha, \beta,\gamma, r_0,R,V$ for $N_1, \bar{N_1},N_5,
\bar {N_5}, n_L,n_R$ defined by
$$\eqalignno{ N_5&={ r_0^2 \over {4 \alpha^{\prime}}} e^{2 \alpha} &(2.7a)\cr
                  \bar {N_5}&={r_0^2 \over {4 \alpha^{\prime}}} e^{-2 \alpha} &(2.7b)\cr
                  N_1&={V r_0^2 \over {4g^2 \alpha^{\prime 3}}} e^{2 \beta} &(2.7c)\cr
                   \bar {N_1}&={V r_0^2 \over {4g^2 \alpha^{\prime 3}}} e^{-2 \beta} &(2.7d)\cr
                   n_L&={R^2V r_0^2 \over {4g^2 \alpha^{\prime 4}}} e^{2 \gamma} &(2.7e)\cr
                   n_R&={R^2V r_0^2 \over {4g^2 \alpha^{\prime 4}}} e^{-2 \gamma} &(2.7f)}$$
In terms of the above numbers, the charges of the black hole are $Q_1=N_1-\bar{N_1}$,
 $Q_5=N_5-\bar{N_5}$, $n=n_L-{n_R}$. The black hole mass is
$$M_{BH}={RV \over {g^2 \alpha^{\prime 3}}}(N_5+ \bar{N_5})+
{R \over {\alpha^{\prime}}} (N_1+\bar{N_1}) + {1 \over R}(n_L+n_R)
\eqno(2.8)$$
The entropy can be written as
$$S=2\pi(\sqrt{N_1}+\sqrt{\bar{N_1}}) (\sqrt{N_5}+\sqrt{\bar{N_5}}) (\sqrt{n_L}+ 
\sqrt{n_R}) \eqno(2.9)$$
The extreme limit is given by $\bar{N_1}=\bar{N_5}=n_R=0$. Small deviations beyond
extremality are given by $\bar{N_1} \sim \bar{N_5} \sim n_R <<N_1 \sim N_5 \sim n_L$.
Note that all three anticharges must become nonzero in order to satisfy the RN nature of the
nonextreme black hole. The RN condition, $\alpha=
\beta=\gamma$ gives two relations which fix two of the moduli
$$V=\left({g^2N_1 \bar{N_1} \over{N_5 \bar{N_5}}} \right)^{1/2} \eqno(2.10)$$ 
and
$$R=\left({n_L n_R \over{N_1 \bar{N_1}}} \right)^{1/4} \eqno(2.11)$$

Our aim in the next section will be to show that the RN black hole mass and entropy
are given by that of an fundamental closed string with a renormalized tension in a five brane
background. We will identify $Q_1$ and $n$ with the net winding and momentum number
of the string whereas $Q_5$ will be the net five brane (winding) number. 
The mass of the black hole will be
given by the sum of the background (five brane) and string masses. The entropy, on the
other hand, will be given simply by that of the string by properly taking the rescaling of the string
tension due to the gravitational effects into account. We will show that this method applies to
both the extreme and nonextreme black holes with any number of charges.

\bigskip
\centerline{\bf 3. Black Holes and Strings with Rescaled Tension}

We begin by examining a five dimensional black hole with only one charge which
was done in refs. [\MAL,\HRS]. We introduce our method of string tension
rescaling for this simple case.  We then generalize this to 
extreme and nonextreme RN black holes with three charges.

{\it 3.1. Black Holes with One Charge}

Consider a five dimensional nonextreme black hole with one NS charge $Q_5=N_5$. 
The metric
for this solution is obtained from eq. (2.1) by taking $\beta=\gamma=0$ and is
given by
$$ds^2=-f^{-2/3} \left(1-{r_0^2 \over r^2} \right)dt^2+f^{1/3} \left[\left(1-
{r_0^2 \over r^2} \right)^{-1}dr^2+r^2d\Omega_3^2 \right] \eqno(3.1)$$
where
$$f=\left(1+{{r_0^2 sinh^2 \alpha} \over r^2} \right)  \eqno(3.2)$$
In terms of the numbers in eq. (2.7) the mass of this black hole is
(here $N_1=\bar{N_1} \sim n_L = n_R <<N_5$)
$$M_{BH}={RV \over {g^2 \alpha^{\prime 3}}}N_5+{R \over \alpha^{\prime}}
(N_1+\bar{N_1}) + {1 \over R}(n_L+n_R)
\eqno(3.3)$$
and the entropy is
$$S=2\pi \sqrt{N_5} (\sqrt{N_1}+\sqrt{\bar{N_1}}) (\sqrt{n_L}+ 
\sqrt{n_R}) \eqno(3.4)$$

We want to show that the nonextreme excitations of this black hole are
given by a closed long elemenatry string with no charge [\MAL,\HRS] so that
$$M_{st}^2={{4N_L} \over \alpha^{\prime}}={{4N_R} \over \alpha^{\prime}} \eqno(3.5)$$
with the oscillator numbers
$$N_{L,R}={1 \over 4}[(\sqrt{N_1}+\sqrt{\bar{N_1}})(\sqrt{n_L}+\sqrt{n_R}) \pm
  (\sqrt{N_1}-\sqrt{\bar{N_1}})(\sqrt{n_L}-\sqrt{n_R})]^2 \eqno(3.6)$$
Note that $N_{L,R}$ above satisfy the level matching condition for any $N_1, \bar{N_1},n_L,n_R$.
The deviation of the black hole
mass from its extreme value is given by the string mass above. The extreme
mass is given simply by that of the five brane background. The entropy is also given solely 
by that of the string since the entropy of the extreme black hole vanishes. 
We remind that for the small neutral deviations from extremality $N_1=\bar{N_1}$, 
 $n_L=n_R$ and $N_1$. Thus for this case
$N_L=N_R=4N_1n_L$ as expected from the level matching conditions for a closed string.
The total mass of the black hole is 
$$M_{BH}={RV \over {g^2 \alpha^{\prime 3}}} N_5 +{2 \over \sqrt{\alpha^{\prime}}}
\sqrt{{4N_1n_L} \over \alpha^{\prime}} \eqno(3.7)$$
This is the same as eq. (3.3) for all deviations from extremality 
(in mass) equal, i.e. $N_1 R/ \alpha^{\prime}=n_L/R$.
Thus, we are able to reproduce the nonextreme black hole mass formula. 
The black hole entropy is more difficult to obtain however. Naively, 
the entropy of the black hole is given by that of the string
$$S=2\pi \sqrt{c \over 6} (\sqrt{N_L}+\sqrt{N_R}) \eqno(3.8)$$
Using the expressions for $N_{L,R}$, we see that this does not give the black hole 
entropy in eq. (3.4). This is not surprising since
the above formula for entropy is true for a free string whereas what we have is a string in a five
brane or black hole background. In the curved background of a black hole
the Rindler energy of an object is given by $dE_R=dM/2 \pi T$. Comparing this with
 $dM=TdS$ we find that
$$S=2\pi E_R \eqno(3.9)$$ 
The Rindler energy is a dimensionless quantity which should be
identified with the (square root of the)
oscillator number of the string and not with its mass. This is also apparent
from eq. (3.9) and the expression for the string entropy.

In the near horizon limit, the metric eq. (3.1) is the Rindler space--time. In this limit
$$r \to r_0, \quad f \to1+sinh^2\alpha=cosh^2\alpha \eqno(3.10)$$
We rescale the metric by
$$r^{\prime}=r \sqrt \lambda, \quad r_0^{\prime}=r_0 \sqrt \lambda \eqno(3.11)$$
Now the charge of the black hole is
$$Q_5=N_5={r_0^2 \over{2 \alpha^{\prime}}}sinh(2\alpha) \simeq {r_0^2 \over
{ \alpha^{\prime}}}cosh^2\alpha \eqno(3.12)$$
where the second equality holds in the near extreme case when $r_0 \to 0, 
\alpha \to \infty$, i.e. $\bar{N_5}=0$.
Then
$$\lambda r_0^{\prime}=r_0 cosh \alpha= \sqrt{Q_5 \alpha^{\prime}} \eqno(3.13)$$
Expanding near the horizon, $r^{\prime}=r_0^{\prime}+y$ the metric becomes
$$ds^2=-\lambda^2{2y\over r_0^{\prime}}dt^2+{r_0^{\prime} \over {2y}}dy^2+ r_0^{\prime2}d\Omega_3^2 \eqno(3.14)$$
The proper distance $\rho$ to the horizon is
$$\rho=\int\sqrt{ r_0^{\prime} \over{2y}} dy=\sqrt{2r_0^{\prime}}\sqrt y \eqno(3.15)$$
Then the coefficient of $dt^2$ becomes
$$g_{00}=-{\rho^2 \over{\lambda^2 r_0^{\prime2}}} \eqno(3.16)$$
One can bring the metric to Rindler form by the rescaling
$$\tau={t \over{\lambda r_0^{\prime}}} \eqno(3.17)$$
where $\tau$ is the Rindler time conjugate to Rindler energy $E_R$ given by
$$E_R=\lambda r_0^{\prime}M=M\sqrt{N_5 \alpha^{\prime}} \eqno(3.18)$$
As a result of eq. (3.9) we identify in the Rindler space--time
$$E_R=\sqrt{c \over 6}\left(\sqrt{N_L^{\prime}}+\sqrt{N_R^{\prime}}\right) \eqno(3.19)$$
Now for a free string with no charge $M^2=4 N_L/  \alpha^{\prime}$. 
We see that, in the five brane background, the string oscillator number $N_L$ is rescaled by a factor of $\sqrt {N_5}$ compared to the free string (for $c=6$); $N_{L,R} \to \sqrt{N_5}
N_{L,R}$. 
On the other hand, the mass of the free string (added to that
of th five brane) gives  the
correct black hole mass. This means that we need to keep the string mass fixed but
rescale $N_{L,R}$. This  can be done if we assume that  
$\alpha^{\prime}$ (or the string tension) is rescaled
simultaneously with $N_{L,R}$ so that the mass remains the same, i.e. $\alpha^{\prime} \to
\alpha^{\prime}_{eff}=N_5 \alpha^{\prime}$. (A similar effect occurs in the D brane picture of this
black hole [\MAL].) 
The entropy of  the string with the rescaled tension is  obtained 
from eqs. (3.19) and (3.6) for $c=6$
$$S=2\pi \sqrt{N_5} (\sqrt{N_1}+\sqrt{\bar{N_1}}) (\sqrt{n_L}+\sqrt{n_R})  \eqno (3.20) 
$$
which is exactly the entropy of the black hole given by eq. (3.4).

We found that a nonextreme five dimensional black hole with one charge can be
described as a long closed string with a rescaled tension in a five brane background. 
Due to the presence of the five brane background, the string tension is rescaled
together with $N_{L,R}$ so that the its mass remains the same. Taking 
this rescaling into account
we find that the string entropy gives exactly the black hole entropy. Here we assume that
that all gravitational effects which are present in the black hole
are summed up in the renormalization of the string tension. Note also that the formula
holds only for $c=6$ whereas for a physical type II string $c=12$. This means that the
string effectively oscillates only in the five brane world--volume and therefore 
has four transverse
directions. We do not know how to justify the above arguments but we will show in 
the next section that they also hold
for extreme and nonextreme RN black holes.

{\it 3.2 Extreme Reissner--Nordstrom Black Holes}

In this section we consider extreme RN black holes with three charges. From eq. (2.8) 
and (2.9) they have  mass  (since $\bar{N_5}=\bar{N_1}=n_R=0$)
$$M_{BH}={RV \over {g^2 \alpha^{\prime 3}}}N_5+
{R \over \alpha^{\prime}}N_1 + {n_L \over R} \eqno(3.21)$$
and entropy 
$$S=2\pi\sqrt{N_1 N_5 n_L} \eqno(3.22)$$
For the RN black hole the three contributions to the mass are equal.
Following the arguments of the previous section, we will describe this black hole as a string with
two charges in a five brane background. Two of the chrages of the black hole, $N_1$ and $n_L$
are the winding and momentum numbers of the string so that the mass of the string becomes
$$M^2_{st}=Q_R^2+{{4N_R} \over \alpha^{\prime}}=Q_L^2+{{4N_L} \over \alpha^{\prime}} \eqno(3.23)$$
with 
$$Q_{R,L}=\left((N_1-\bar{N_1}){R \over \alpha^{\prime}} \pm {(n_L-n_R) \over R} \right) \eqno(3.24)$$
Using the definitions of $N_{L,R}$ from eq. (3.6) we find that for the extreme black hole
$N_R=0$ and $N_L=4N_1 n_L$.
Thus we find that the string is in a BPS state. This is not surprising since we are considering 
an extreme RN black hole which is described by a BPS string state on a BPS background 
(the five brane). The mass of the five brane with the string is
$$M={RV \over {g^2 \alpha^{\prime 3}}}N_5+{4N_1 n_L \over \alpha^{\prime}} \eqno(3.25)$$
which is equal to that of the black hole. As before, since we are considering
not a free string but one in a five brane backgound the entropy is again given by
$S=2 \pi E_R$. The redshift factor due to the five brane is 
again $N_{L,R} \to \sqrt{N_5} N_{L,R}$  and we find (for $c=6$)
$$S=2\pi \sqrt{N_L^{\prime}}=2 \pi \sqrt{N_5 N_1 n_L}  \eqno(3.26)$$
which is the correct entropy for the extreme black hole.
Clearly, after the rescaling of the string tension, the smallest excitation of the extreme
black hole given by $\delta N_R=1$ corresponds to an energy $\delta M \sim 1/N_5 N_1 R$
as required.
This is analogous to the statement that the string becomes one long string of length 
$N_5 N_1 R$ in the D brane picture[\MS].

In this case, the division of the black hole into a background and a string is
not clear since the string weighs twice as much as the background. (For the one 
charge case the mass of the string which is the deviation from extremality is much
smaller than the background mass, i.e. the extreme mass.) 
On the other hand, a string can only carry two
charges which makes the  five brane background necessary if we want to describe
a black hole with three charges. In addition, only a background
with one charge (such as the five brane) has a Rindler space near its horizon so 
that the above method can be applied. The case with two charges is a straightforward
generalization of the above and we will not discuss it.

{\it 3.3 Nonextreme Reissner--Nordstrom Black Holes}

We now consider slightly nonextreme RN black holes with mass
$$
M_{BH}={RV \over {g^2 \alpha^{\prime 3}}}(N_5+ \bar{N_5})+
{R  \over \alpha^{\prime}}(N_1+\bar{N_1}) + {1 \over R}(n_L+n_R)   
\eqno (3.27)
$$
and entropy
$$S=2 \pi 
(\sqrt{N_1}+\sqrt{\bar{N_1}}) (\sqrt{N_5}+\sqrt{\bar{N_5}}) (\sqrt{n_L}+ 
\sqrt{n_R}) 
\eqno(3.28)
$$
with $N_1 \sim N_5 \sim n_L >>\bar{N_1} \sim \bar {N_5} \sim n_R$. Note 
that the deviation from
extremality is for the three charges simultaneously due to the RN nature of the black hole.
Since the black hole is
nonextreme it is described by a non BPS string state ($N_R \not=0$) in a
nonextreme background (five branes and anti five branes). The string mass is given by
$$M^2_{st}=Q_R^2+{{4N_R} \over \alpha^{\prime}}=Q_L^2+ {{4 N_L} \over \alpha^{\prime}} \eqno(3.29)$$
with 
$$
Q_{R,L}=\left((N_1-\bar{N_1}){R \over \alpha^{\prime}} \pm {(n_L-n_R) \over R} \right) \eqno(3.30)
$$
with $N_{L,R}$ given by eq. (3.6). Using the condition for the extreme RN case (which holds
approximately for the nonextreme case) 
$${{N_1 R} \over \alpha^{\prime}} \simeq {n_L \over R}  \eqno(3.31)$$
we find that the string mass is
$$M_{st}=\sqrt{{4N_1 n_L} \over \alpha^{\prime}}+{{8 N_1 n_R} \over \alpha^{\prime}} \eqno(3.32)$$
For anti charges much smaller than the charges we find that 
$$M_{st} \simeq {{2 N_1  R} \over \alpha^{\prime}} +{2 n_R \over R} \eqno(3.33)$$
Adding this to the mass of the background five and anti five branes we find
$$M={RV \over {g^2 \alpha^{\prime 3}}}(N_5+ \bar{N_5})+{{2R} \over  \alpha^{\prime}}
N_1  + {2 \bar{N_1} \over R} \eqno(3.34)$$
which is the black hole mass. 

In order to find the entropy, we have to calculate the Rindler energy of the above string
$E_R$ in the background of
$N_5$ five branes and $\bar {N_5}$ anti five branes with $N_5>>\bar {N_5}$. Once again
we find 
$$E_R=M{r_0 \over {2 \alpha^{\prime}}} sinh (2 \alpha) \sim  {Mr_0 \over \alpha^{\prime}}
cosh \alpha  \eqno (3.35)$$
in the near extreme limit. But now,
$$Q_5=N_5-\bar{N_5}={r_0^2 \over \alpha^{\prime}} sinh^2 (2 \alpha) \eqno(3.36)$$
so that the oscillator numbers and the string tension are rescaled by the factor $\sqrt{N_5+\bar{N_5}}$. The entropy of the string is 
$$S=2 \pi E_R=2 \pi (\sqrt{N_L^{\prime}}+\sqrt{N_R^{\prime}}) \eqno(3.37)$$
where $N_{L,R}^{\prime}=\sqrt{N_5+\bar{N_5}} N_{L,R}$. Using the definition of $N_{L,R}$
in eq. (3.6) we find the entropy of the string in the background
$$S=2\pi(\sqrt{N_1}+\sqrt{\bar{N_1}}) (\sqrt{N_5}+\sqrt{\bar{N_5}}) (\sqrt{n_L}+ 
\sqrt{n_R}) \eqno(3.38)$$
which is exactly the entropy of the nonextreme black hole. The rescaling factor of the tension
$\sqrt{N_5+\bar{N_5}}$ is exactly the factor needed to convert the string entropy into the
black hole entropy.

\bigskip
\centerline{\bf 4. Hawking Radiation of Scalars}

In the previous section, we saw that the RN black hole is described by a long closed string
with a rescaled tension. In this section, we calculate the rate of emission of a low energy
scalar from a slightly nonextremal string in a five brane and find that it matches the Hawking radiation
expected from the corresponding black hole. Since $N_5=1$ the string tension is not
rescaled in this case. 
The presence of the five brane background does not affect scalar emission
from the string except for confining the string oscillations into the brane world--volume as 
is evident  
from the fact that $c=6$. 
This calculation which is performed for small coupling is analogous to that of ref. [\DM] in the D brane picture.

Consider a non BPS state of the string with $N_L>>N_R \not =0$. From the formulae for the
mass and entropy we find the left and right temperatures
$$\beta_{L,R}={M \pi \over 4} \sqrt{c \over 6} {1 \over \sqrt{N_{L,R}}} \eqno(4.1)$$
We also define the world--sheet temparatures[\HKRS]
$$\beta^*_{L,R}={dS \over dN_{L,R}}= \sqrt{\pi^2 c \over{6N_{L,R}}} \eqno(4.2)$$
$\beta^*$ is a dimensionless parameter which is chosen to fix the average value of the string
oscillator numbers $N_{L,R}$. The statistical emsemble which describes the thermal initial
state of the string described by the density matrix
$$\rho=Z^{-1}exp(-\beta_LN_L-\beta_RN_R) \eqno(4.3)$$
where $Z$ defined so that $Tr \rho=1$.
We can now  calculate the rate of emission of a scalar, $g_{56}$ component of the
graviton. (Here $5$ and $6$ refer to two of the transverse directions inside the five brane.) 
The vertex operator for this scalar is
$$V(k)=\int{4 \sqrt2 \over \pi}[\partial_+X^5 \partial_-X^6+ fermion \quad terms]e^{ikX} d\sigma
\eqno(4.4)$$
where the derivatives refer to world--sheet light--cone coordinates. If the momentum of the
emitted scalar $k$ is much smaller than the string length $\ell_s$, the exponential factor
$e^{ikX}$ and the fermionic terms in the vertex operator can be neglected. The matrix element
for the decay of an initial string state $|i \rangle$ at rest to a final state $|f \rangle$ by
emitting a scalar $g_{56}$ is[\HKRS]
$${\cal M}={4 \sqrt2 \over \pi} \langle i| \int \sum_{n,m} \alpha_n^5 \tilde{\alpha_m}^6
e^{-2i(n-m) \sigma} d \sigma |f \rangle \delta^4(p_i+k-p_f) \eqno(4.5)$$
Using the mass formula for the string 
$$M^2=Q_R+8N_R=Q_L^2+8N_L \eqno(4.6)$$
(here $\alpha^{\prime}=1/2$ since $N_5=1$)
and defining $\delta N=N_R=n$ and $\delta M=\omega$
we find
$$\omega={{4n} \over M} \eqno(4.7)$$
Integrating over the world--sheet coordinate $\sigma$ we obtain the condition $n=m$. 
This means
that two oppositely moving oscillators with energy $n$ collide to give a space--time scalar
$g_{56}$ with energy $\omega$. In order to obtain the decay rate we square the amplitude, 
average over the initial thermal distribution and sum over the final states
$$|{\cal M}^{\prime}|^2Tr \rho \sum_f|{\cal M}|^2=32\kappa^2 n^2{1 \over (e^{\beta^*_Ln}-1)} {1 \over (e^{\beta^*_Rn}-1)}
{2 \over (2M)^2} {M \over 4}\eqno(4.8)$$
where the factor of two is due to the two polarizations $g_{56}$ and $g_{65}$, the factors of
$1/2M$ are for the relativistic normalization of the initial and final states and $M/4$ is the density of the resonances of the string.
Note that since $N_L>>N_R$,  $\beta^*_R>>\beta^*_L$. Also from eq. (4.2) we see that
$\beta^*_L n<<1$ so that it can be expanded in a power series. Using the expression for
$\beta^*_L$ we get
$$|{\cal M}^{\prime}|^2={\kappa^2 \omega \over \pi} \sqrt{6 N_L \over c} {1 \over (e^{\beta^*_Rn}-1)} \eqno(4.9)$$
The decay rate is given by Fermi's golden rule
$$\Gamma=|{\cal M}^{\prime}|^2 2\pi {1 \over {2 \omega}} 
{d^4\omega \over (2\pi)^4}  \eqno(4.10)$$
Substituting the amplitude ${\cal M}$  we find
$$\Gamma={6 \kappa^2 \over c} {S \over {2 \pi}} 
{1 \over (e^{\beta^*_Rn}-1)}  {d^4\omega \over (2\pi)^4} \eqno(4.11)$$
where $S=2\pi \sqrt{cN_L/6}$ is the entropy of the extreme black hole. For $c=6$ and using the relation $\kappa^2=8\pi G_N$ we find
$$\Gamma=4G_N S {1 \over (e^{\beta^*_Rn}-1)} {d^4\omega \over (2\pi)^4} \eqno(4.12)$$
Using the Bekenstein--Hawking formula $A_H=4 G_N S$
 we find that the rate of emission of the scalar $g_{56}$ from the string is given by the area
of the black hole horizon as required. Scalar emission is also thermal due to the thermal factor in
eq. (4.12). Note that
$$\beta^*_R n=\beta_R \omega \eqno(4.13)$$
where $\beta_R=1/T_H$ is the Hawking temparature. Therefore, the slightly nonextreme
string radiates thermally 
at exactly the Hawking temperature. This gives further evidence to our description of the
RN black hole as a closed string with a rescaled tension in a five brane background.
Since $c=6$ the string can oscillate only in the five brane world--volume. As a result,
emission of vector bosons and gravitons are not possible at the lowest order. These
are emitted from the string by higher order world--sheet processes involving fermionic
vertex operators. As in the D brane picture, it is interesting that the emission from a
weakly coupled string reproduces the Hawking radiation from a black hole which
corresponds to the large coupling region.

\bigskip
{\bf 5. Conclusions and Discussion}

We have shown that extreme and nonextreme NS charged RN black holes in five dimensions
can be
described by fundamental strings with two charges in a magnetic five brane.
Due to the five brane background, the string oscillator number
and tension are rescaled whereas its mass remains fixed. In addition, the string 
oscillations are confined to the five brane world--volume.
The extreme and nonextreme
black holes are described by BPS and non BPS states of the string respectively.
The mass of the black hole is given by the sum of masses of the string and the five brane.
The entropy, on the other hand, is given by only that of the string with the tension
rescaling taken into account. In addition, the Hawking radiation of a low energy scalar 
from the black hole can be described by the emission of the scalar from such a string
with a rescaled tension and confined into the five brane.

The picture we advocated above relies on a number of assumptions which we are 
not able to justify.
First, we assume that all gravitational effects in the black hole are summed up by the
rescaling of the string oscillator number and tension. We do not know how to explain this
which probably requires a deeper understanding of strings in Rindler space--times.
Second, the string mass remains fixed for both BPS and non BPS states in the
five brane background. This is expected for the BPS states but not for the non BPS
ones. We must note however that similar assumptions are made in the D brane
picture of RR charged black holes. Third, we neglect the possible gravitational
effects of the string mass on the string itself. This is difficult to understand for RN
black holes for which the string is twice as heavy as the five brane background.
Nevertheless, we see that these assumptions lead us to a microscopic description of 
extreme and nonextreme NS charged RN black holes in five dimensions..

The NS charged RN black holes are of course connected to RR charged ones by S
duality of the type IIB string theory. Thus, we expect the above picture to be connected
to the microscopic description of black holes by D branes. In fact, under S duality, the
D five branes, D strings and momentum of ref. [\SV] are transformed into magnetic five
branes and fundamental strings of the above description. In both cases the string
is confined into the world--volume of the five brane. Also in both cases, masses 
of the black hole constituents are simply additive even when the black hole is nonextreme.
In the D brane picture, the presence of the five branes effectively multiply
the number of degrees of freedom by forming a long string. In the above picture, the
same happens by the rescaling the oscillator number of the string due to the
gravitational five brane background. Thus, we see that the two descriptions of black 
holes are very similar as expected from S duality. There are however, a number of differences
which are important. First, the above picture holds for both type IIA and IIB string 
theories whereas the D brane picture (for the five dimensional black holes) holds only
for the latter. Second, the D brane description is in flat space--time whereas the one
above is in the curved space--time of the background. 
Third, in this picture, we are able to compute the nonextreme black hole
entropy when all three deviations from extremality (in mass) 
are nonzero simultaneously. This is
not the case for the D brane picture where one does this one at a time and then
resorts to U duality. 

It seems that our method also works for four dimensional and/or rotating RN black
holes in a straightforward fashion.  It would be interesting to examine
emission of particles with spin and scalars
from strings in more than one five brane. These issues
are currently under investigation and will be reported elsewhere.

\bigskip
\centerline{\bf Acknowledgments}
This work originated from discussions with L. Susskind. We would like to thank him
for very useful discussions and for reading the manuscript.

\vfill
\eject

\refout
\vfill
\eject

\end